\documentclass[prb,twocolumn,showpacs,preprintnumbers,amsmath,amssymb,showkeys,superscriptaddress]{revtex4}

\usepackage{graphicx}
\usepackage{dcolumn}
\usepackage{bm}

\begin{document}


\title{Strain and correlation of self-organized $Ge_{1-x}Mn_x$ nanocolumns embedded in Ge (001)}

\author{S. Tardif}
 \affiliation{Institut N\'eel, CNRS and Universit\'e Joseph Fourier, 25 rue des Martyrs, BP166, F-38042 Grenoble, France}
 \affiliation{CEA-UJF, INAC, SP2M, 38054 Grenoble, France}
\author{V. Favre-Nicolin}
 \affiliation{CEA-UJF, INAC, SP2M, 38054 Grenoble, France}
 \affiliation{Universit\'e Joseph Fourier, Grenoble, France}
\author{F. Lan\c con}
 \affiliation{CEA-UJF, INAC, SP2M, 38054 Grenoble, France}
\author{E. Arras}
 \affiliation{CEA-UJF, INAC, SP2M, 38054 Grenoble, France}
\author{M. Jamet}
 \affiliation{CEA-UJF, INAC, SP2M, 38054 Grenoble, France}
\author{A Barski}
 \affiliation{CEA-UJF, INAC, SP2M, 38054 Grenoble, France}
\author{C. Porret}
 \affiliation{CEA-UJF, INAC, SP2M, 38054 Grenoble, France}
\author{P. Bayle-Guillemaud}
 \affiliation{CEA-UJF, INAC, SP2M, 38054 Grenoble, France}
\author{P. Pochet}
 \affiliation{CEA-UJF, INAC, SP2M, 38054 Grenoble, France}
\author{T. Devillers}
 \altaffiliation{Present address: Institut f\"ur Halbleiter-und-Festk\"orperphysik Johannes Kepler Universit\"at Altenbergerstr. 69, A-4040, Linz, Austria}
 \affiliation{CEA-UJF, INAC, SP2M, 38054 Grenoble, France}
\author{M. Rovezzi}
 \altaffiliation{Present address: Institut f\"ur Halbleiter-und-Festk\"orperphysik Johannes Kepler Universit\"at Altenbergerstr. 69, A-4040, Linz, Austria}
 \affiliation{Consiglio Nazionale delle Ricerche, GILDA CRG, c/o ESRF 6 Rue Jules Horowitz F-38043 Grenoble, France}

\date{\today}

\begin{abstract}
We report on the structural properties of $Ge_{1-x}Mn_x$ layers grown by molecular beam epitaxy. In these layers, nanocolumns with a high Mn content are embedded in an almost-pure Ge matrix. We have used grazing-incidence X-ray scattering, atomic force and transmission electron microscopy to study the structural properties of the columns. We demonstrate how the elastic deformation of the matrix (as calculated using atomistic simulations) around the columns, as well as the average inter-column distance can account for the shape of the diffusion around Bragg peaks.

\end{abstract}

\pacs{75.50.Pp, 61.46.-w,61.05.cp, 75.75.-c}
\keywords{ferromagnetic semiconductor, grazing incidence X-ray scattering, atomistic simulations}

\maketitle

\section{\label{sec:intro} Introduction}

Ferromagnetic semiconductors have been extensively studied over the last decade as they are considered as the material solution for the needs of the spin-based electronics. Among the possible candidate systems, germanium-based ones have the advantage of being compatible with existing main-stream silicon technology. In particular, early results on the control of the ferromagnetism ordering in gated germanium-manganese structures using an electric field\cite{park_group-iv_2002} have spurred many studies, focusing either on the diluted \cite{cho_ferromagnetism_2002,picozzi_x-ray_2005,li_ferromagnetic_2005,morresi_magnetic_2006,ottaviano_direct_2006,de_padova_morphological_2006,chen_magnetic_2007,zeng_optimal_2008} or the heterogeneous \cite{park_magnetoresistance_2001,kang_spatial_2005,sugahara_precipitation_2005,ahlers_ferromagnetic_2006,bihler_structural_2006,bougeard_clustering_2006,jamet_high-curie-temperature_2006,ottaviano_nanometer-scale_2006,biegger_intrinsic_2007,devillers_structure_2007,li_dopant_2007,de_padova_mn0.06_2008,hol_diffuse_2008,kazakova_effect_2008,wang_direct_2008,xiu_electric-field-controlled_2010} aspects of the GeMn system. Considering epitaxially grown GeMn systems, several groups have observed the self-assembly of Mn-rich, columnar-like nano-objects embedded in a Ge matrix\cite{jamet_high-curie-temperature_2006,bougeard_clustering_2006,ahlers_magnetic_2006,devillers_structure_2007,li_dopant_2007}. Curie temperatures above room temperature have been observed by Jamet \textit{et al.} in these nanocolumns\cite{jamet_high-curie-temperature_2006}, and also more recently by Cho \textit{et al.} in GeMn nanowires\cite{cho_ferromagnetic_2008}, by Zeng \textit{et al.} in homogeneous Mn-doped Ge thin films with $x_{Mn}=0.25$\%\cite{zeng_optimal_2008} and by Xiu \textit{et al.} in Mn$_{0.05}$Ge$_{0.95}$ quantum dots\cite{xiu_electric-field-controlled_2010}, thus reinforcing the interest in this system.  

\begin{figure}
	\centering
	\includegraphics[width=0.48\textwidth]{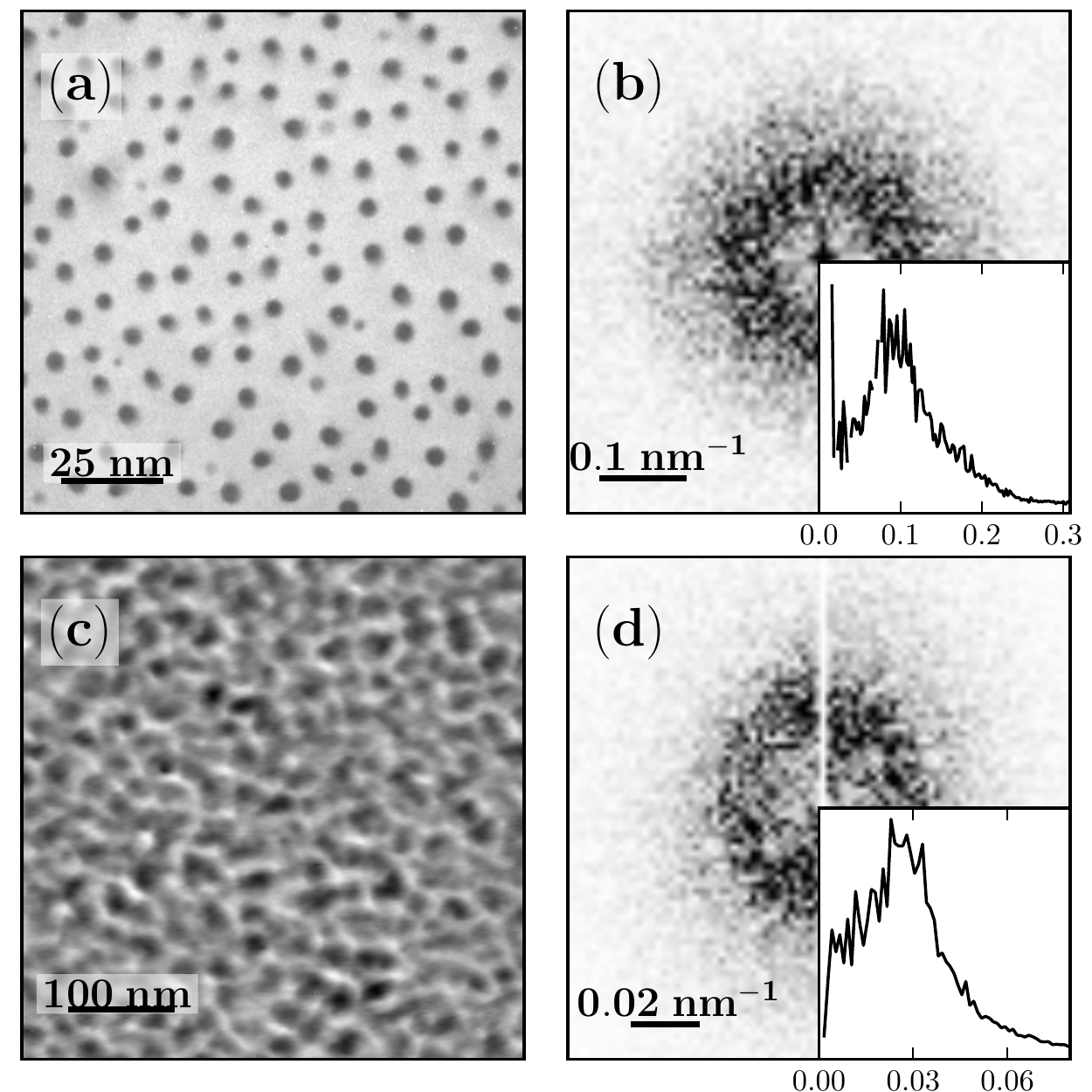}
	\caption{\label{fig:FFT} $(a)$ Plane-view TEM micrograph, $(c)$ surface topography measured by AFM, and $(b),(d)$ the central zones of the corresponding Fourier transforms (FT), each with an inset showing the radial distribution of the FT amplitude. All color scales are linear, with the full z-scale for the AFM image corresponding to 2.6 nm. In each FT image a more intense ring can be observed, with correlation distances equal to $\sim 10$ and $\sim 40$ nm, respectively in the TEM (inter-column average spacing) and the AFM (surface roughness) images.}
\end{figure}

The low-temperature molecular beam epitaxy of Ge and Mn allows for a high concentration of Mn (up to 10\%) to be incorporated in the Ge matrix. Due to a spinodal decomposition occurring at the early stages of the growth, Mn-rich regions form. Depending on the growth parameters (substrate temperature, growth rate, manganese concentration), these inclusions can be spherical,\cite{ahlers_magnetic_2006} cigar-shaped\cite{ahlers_magnetic_2006} or columnar and extending throughout the GeMn layer\cite{jamet_high-curie-temperature_2006,devillers_structure_2007,li_dopant_2007}. If the growth is performed at higher temperature or if the film is annealed, more stable phases can be seen, such as Mn$_5$Ge$_2$, Mn$_5$Ge$_3$ or Mn$_{11}$Ge$_8$ clusters.\cite{passacantando_growth_2006,bihler_structural_2006,li_magnetic_2006,ahlers_magnetic_2006,morresi_formation_2006,ayoub_structural_2006,de_padova_mn0.06_2008,wang_direct_2008,ahlers_stefan_magnetic_2008,devillers_etude_2008,lechner_self-assembled_2009,jain_abhinav_investigation_2010} Many studies have shed light on the magnetic and electronic properties of the GeMn nanocolumns and inclusions.\cite{jamet_high-curie-temperature_2006,bougeard_clustering_2006,ahlers_magnetic_2006,devillers_structure_2007,devillers_structural_2007, devillers_etude_2008,ahlers_stefan_magnetic_2008,bougeard_ge1-xmnx_2009,ahlers_comparison_2009} In terms of structural properties, the relationship between the GeMn layer and the Mn$_5$Ge$_3$ clusters is now well known.\cite{hol_diffuse_2008,lechner_self-assembled_2009,jain_abhinav_investigation_2010} Although information about the \textit{local} environment around Mn atoms has already been reported,\cite{rovezzi_atomic_2008} the understanding of the structural properties of continuous GeMn nanocolumns and their surrounding matrix is not complete.

In this work we report on the strain and correlations of GeMn nanocolumns embedded in a Ge matrix, as determined by grazing-incidence X-ray scattering (GIXS) techniques at the European Synchrotron Radiation Facility (ESRF), as well as elastic and atomistic simulations. While nanocolumns with diameters ranging from 1.5 to 6 nm can be synthesized, this article focuses on samples with larger nanocolumns, where the inner part of the columns are either disordered or amorphous\cite{devillers_structure_2007}.

The article is organized as follows: in part II we will present the experimental techniques used for the synthesis and the characterization (atomic force and transmission electron microscopy as well as X-ray scattering); in part III we will show, using elastic and atomistic simulations, how the GIXS maps can be interpreted in terms of strain and correlations of the GeMn nanocolumns and the surrounding Ge matrix.

\section{\label{sec:exp} Experiments} 
\subsection{Sample preparation}

The (Ge,Mn) thin films were obtained using solid sources molecular beam epitaxy on epi-ready Ge(001) wafers. First, a 40 nm-thick Ge buffer was grown at 250~$^\circ$C after thermal desorption of the native oxide. Then Ge and Mn atoms were co-evaporated in order to get a 60 to 80 nm-thick(Ge,Mn) layer with an overall Mn concentration ranging from $\sim$~6~$\%$ to $\sim$~10~$\%$, depending on the relative ratio of the Mn and Ge evaporation flux. The growth temperature was set to $\sim$100~$^\circ$C and the growth rate was $\sim$~0.2~$\AA.s^{-1}$.

These growth parameters allow for a two-dimensional spinodal decomposition to take place at the early stages of the growth. Due to the layer-by-layer growth mode, the resulting (Ge,Mn) films are heterogeneous and made of columnar structures (nanocolumns) embedded in a matrix. In these conditions, nanocolumns have diameters ranging from 3 to 5 nm and densities between 10~000 and 20~000~$\mu m^{-2}$. The chemical analysis in a transmission electron microscope has shown that the composition in the nanocolumns is close to Ge$_2$Mn while the surrounding matrix is almost pure Ge ($<$1$\%$ Mn).

It was also observed that samples not kept in an dry, oxygen-free environment would slowly oxidize over time, with a full oxidation occurring naturally after $\sim$ 2 years of exposure. The effects of partial oxidation on the magnetic properties of the GeMn nanocolumns have already been reported.\cite{tardif_samuel_exchange_2010}

\begin{figure}
	\centering
	\includegraphics[width=0.5\textwidth]{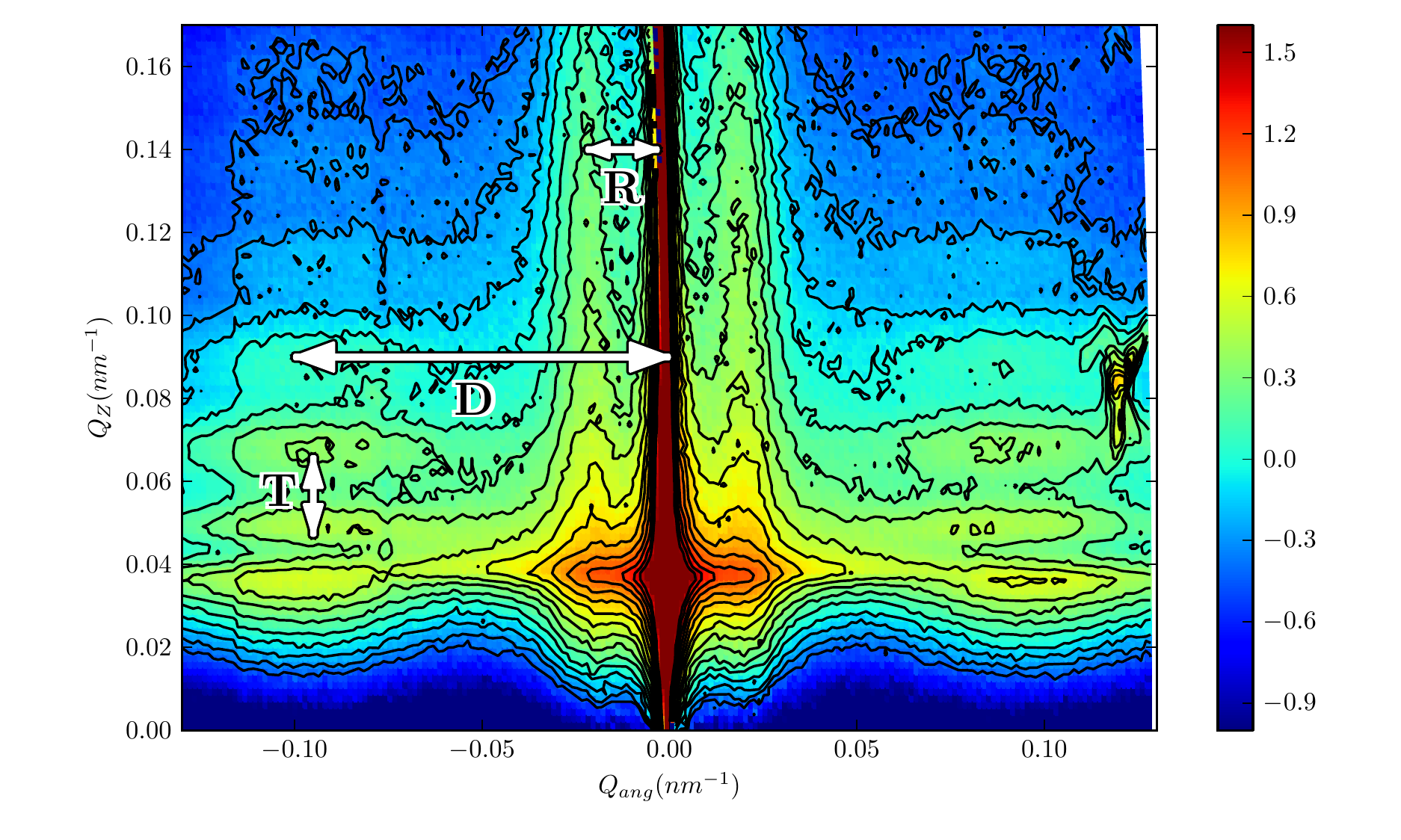}
	\caption{\label{fig:GISAXS} (color online) Bragg-GISAXS (see text for details) map measured by an angular scan around the in-plane (220) Bragg reflection, using a Vantec position-sensitive detector, at E=6.5 keV and $\alpha_i = 0.4^\circ$ (intensity plotted on a logarithmic scale, with 8 contour levels per order of magnitude). $Q_{ang}$ and $Q_z$ are respectively parallel and perpendicular to the surface. Three finite distances effects can be seen, symmetrically about the $Q_{ang} = 0$ axis~: $(i)$ a correlation distances of about 9 to 15nm (marker 'D'), corresponding to the average distance between columns, $(ii)$ the oscillations (marker 'T') along $Q_z$ (dot-dashed line) are due to the finite thickness of the sample (60 nm), and $(iii)$ the two intense streaks (marker 'R') at very small $Q_{ang}$ have been attributed to the surface roughness.}
\end{figure}

\subsection{Transmission electron microscopy}

The samples were observed in cross-section and plane view by transmission electron microscopy (TEM) using a JEOL 4000EX microscope with an acceleration voltage of 400 kV. Standard preparation, including mechanical polishing and argon ion milling, was performed prior to the observations. Preparation for plane views also included wet etching using a H$_{3}$PO$_{4}$-H$_{2}$O$_{2}$ solution.

The nanocolumns can be seen in plane-view TEM micrographs (Fig.\ref{fig:FFT}a) as circular shaped objects with a different contrast from their surroundings. The nanocolumns induce strain in the matrix, as evidenced by the observation in high resolution TEM of a dark ring around them.\cite{jamet_high-curie-temperature_2006,devillers_structure_2007} In some cases, the presence of a pair of dislocation\cite{jamet_high-curie-temperature_2006} allowed for an estimation of the in-plane strain of about 4$\%$. No such strain could be observed in cross-section TEM, where chemical contrast has shown that the columns are continuous, extending from the buffer layer to the surface.

A fast-Fourier transform (FFT) analysis of the plane views reveals a higher intensity ring, as shown in Fig.\ref{fig:FFT}(b). Angular integration of the FFT as well as direct computation of the radial distribution function of the nanocolumns give an in-plane correlation distance between first neighbors of $\approx8\ nm$.

\subsection{Atomic Force Microscopy}

The surface of the samples was studied using atomic force microscopy (AFM). The surface presents a root-mean-square roughness r$_{RMS}$~=~0.8$\pm$0.3 nm, as measured over 500$\times$500 nm$^{2}$ areas (Fig.\ref{fig:FFT}c) . A FFT analysis, shown in Fig.\ref{fig:FFT}d, similar to that performed on the plane view TEM micrographs, exhibits a higher intensity ring at distances of about 40 nm, corresponding to a surface roughness correlation length.

\begin{figure} 
	\centering
	\includegraphics[width=0.42\textwidth]{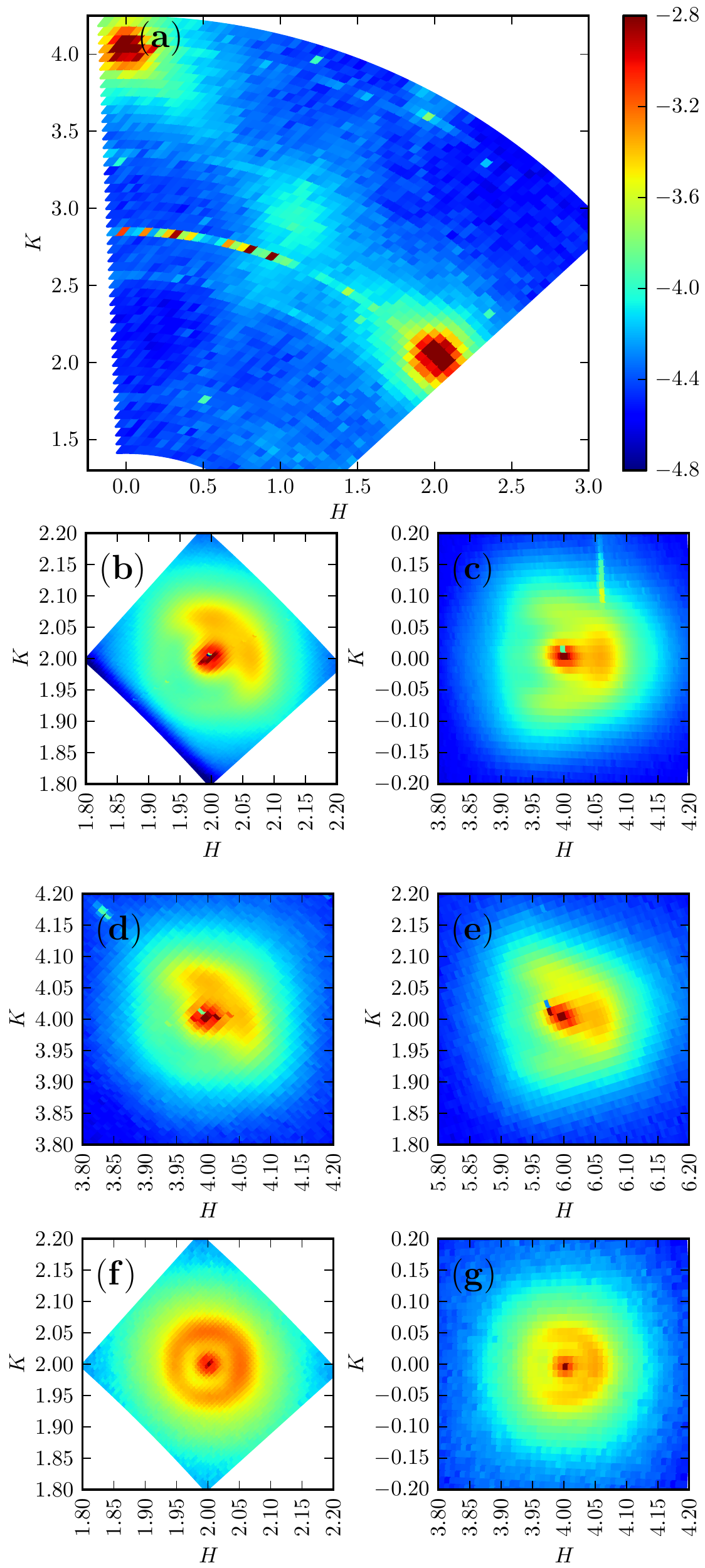}
	\caption{\label{fig:CartoDiffraction} (color online) $(a)$ Grazing-incidence reciprocal space map (RSM) of the measured intensity in the hk0 plane - this map features almost only the $(220)$ and $(400)$ Ge Bragg peaks with a weak diffuse line between them. $(b-e)$ high-resolution RSM around the $(220)$, $(400)$, $(440)$ and $(620)$ reflections. Around each Bragg peak an asymmetric ring can be seen at $\approx0.07$ reciprocal lattice units (r.l.u.): its positions is given by the average inter-column distance while the asymmetry of the intensity is due to the compression of the matrix surrounding each column. In (f-g) are shown the RSM around reflections $(220)$ and $(400)$, for the same type of sample as for Fig.\ref{fig:GISAXS}, but after oxidization (due to aging and exposure to air) of the sample: the correlation ring is almost circular due to the decrease of the strain of the Ge matrix around the oxidized columns. The intensity is represented as a logarithmic color scale, the full scale corresponding to 3 orders of magnitude for all plots.}
\end{figure}

\subsection{Grazing incidence X-ray scattering}

In order to get a better understanding of the correlations between the nanocolumns and their effects on the strain in the matrix, we performed X-ray scattering measurements on beamlines ID01, BM32 and BM02 at the ESRF. We used energies around the Mn K edge (6.539 keV) and the Ge K edge (11.103 keV). In order to enhance the signal from the GeMn layer, we used a grazing incidence geometry, with the incident angle tuned around the total reflection angle ($\alpha _{c}$ = 0.38$^{\circ}$ at the Mn K edge and 0.19$^{\circ}$ at the Ge K edge) in order to probe depths between 10 nm and 0.5~$\mu$m. 

The scattered intensity was measured using both Grazing Incidence Small Angle X-ray Scattering (GISAXS)\cite{renaud_probing_2009} as well as wide angle geometry (close to Bragg reflections). GISAXS measurements are sensitive to the average shape and positional correlations between objects at or close to the surface, but in our case contrast in the average electronic density between the columns and the surrounding matrix is very low, which prevents the observation of the nanocolumn shape and distance correlation using this method.

However a GISAXS-type analysis could still be performed around Bragg reflections: as the detector used for wide-angle measurements was a Vantec position-sensitive 1D detector oriented perpendicularly to the sample plane, we measured the intensity as a function of the exit angle. In the case of an angular scan around a $(220)$ reflection, as shown in Fig.\ref{fig:GISAXS}, the 2D data presents features similar to classical GISAXS: the `enhanced contrast' is simply due to the sensitivity of GIXS near a Bragg reflection to any deformation (e.g. induced by the columns) in the layer.

Two peaks can be seen in the `Bragg-GISAXS' image: one (with $d\approx40\ nm$) corresponding to the surface roughness correlation (as shown in Fig.\ref{fig:FFT}d, and already reported by Hol\`y \textit{et al.}\cite{hol_diffuse_2008}) and another (with $d\approx8\ nm$) corresponds to the average distance between neighboring columns, as shown in Fig.\ref{fig:FFT}b. Moreover, the latter peak exhibits fringes with a period corresponding to the thickness of the GeMn layer (60~nm), which is only possible because the columns are continuous over the whole thickness of the layer.\footnote{In the small-angle regime, a continuous \textit{average density} in the layer is enough to create oscillation fringes - however in `Bragg-GISAXS', such fringes can only be present if the \textit{lattice} is continuous throughout the entire layer.}

In the large in-plane grazing-incidence X-ray scattering maps, shown in Fig.\ref{fig:CartoDiffraction}a, the most important feature is the absence of significant contributions far from the Bragg reflections of the standard diamond lattice. This is true for samples which are the focus of this article, with nanocolumns featuring a large ($\ge 4$ nm) average diameter, and either a disordered or an amorphous inner structure.\cite{devillers_structure_2007}

A weak diffuse streak is also visible between the $(220)$ and $(040)$ reflections, with a faint diffuse $(130)$ Bragg peak - these could either be due to diffuse scattering from the rough surface, or be related to the inner structure of the nanocolumns (which is partially crystalline in the samples considered here) - this will not be the focus of this article, which mostly deals with the strain of the surrounding matrix and the correlations between nanocolumns. We will now focus on these characteristics, which can be analyzed using the shape of the diffusion around the germanium Bragg peaks. 

\section{\label{sec:InterpretXRayScattMaps} Interpretation of the X-ray scattering maps near Bragg reflections}

A specific diffusion is observed around all Bragg peaks, as can be seen in Fig.\ref{fig:CartoDiffraction}(b-e): this type of scattering pattern is due to the deformation of the diamond lattice, since the nanocolumns have a larger average lattice parameter  \cite{jamet_high-curie-temperature_2006,devillers_structure_2007} than that of bulk Ge. 

In Fig.\ref{fig:CartoDiffraction} maps are shown for two different samples: Fig.\ref{fig:CartoDiffraction}(a-e) correspond to the same sample with a strong asymmetry of the correlation ring, whereas Fig.\ref{fig:CartoDiffraction}(f-g) correspond to a similar sample measured after aging, and where the asymmetry is no longer present. This aging effect has alreay been observed quantitatively using SQUID magnetometry,\cite{tardif_samuel_exchange_2010}, and has been linked to the sample oxidation. In the case of a very old sample ($\approx$ 3 years in this case), it is expected that the columns are fully oxidized. Although no direct observation  (using TEM) of the nanocolumns morphology has been made, the strong decrease of the asymmetry can be related to the relaxation of the matrix surrounding the nanocolumns, as we will now study.

In the case of GeMn nanocolumns it was shown \cite{jamet_high-curie-temperature_2006} that the surrounding matrix is compressed. Therefore a Huang-type diffusion \cite{huang_x-ray_1947,dederichs_diffuse_1971,pietsch_high-resolution_2004} could be expected - however the decrease of the intensity around a given reflection (\textit{i.e.} when following the intensity in a radial direction from the center of the Bragg position) is not monotonous, and features a maximum around a ring located at $\Delta (H\ or\ K)\approx0.07\ (r.l.u.)$, \textit{i.e.} corresponding to a distance of ~$\approx8\ nm$ in real space. This distance corresponds approximately to the average inter-column distance, and therefore the location of the ring is due to the inter-column distance correlation, while the polarization of this ring (the intensity is larger on the high-q side, and also larger along $[100]$ and $[010]$ directions) is due to the deformation field of the matrix, as we will now show using elastic and atomistic simulations.

\subsection{\label{sec:AnalyticalModel}Analytical model of the elastic deformations}

The scattering from the GeMn layer can be calculated as the difference between the scattering of the distorted and perfect (\textit{i.e.} without deformation or columns) lattices \cite{pietsch_high-resolution_2004}- the advantage of this method is that it avoids most oscillation fringes due to the finite size of the simulated lattice, and also that it allows the superposition of the contributions from several insertions (columns in our case). The only approximation is that the sharp Bragg peak will not be obtained by this calculation.

\begin{figure} 
	\centering
	\includegraphics[width=0.48\textwidth]{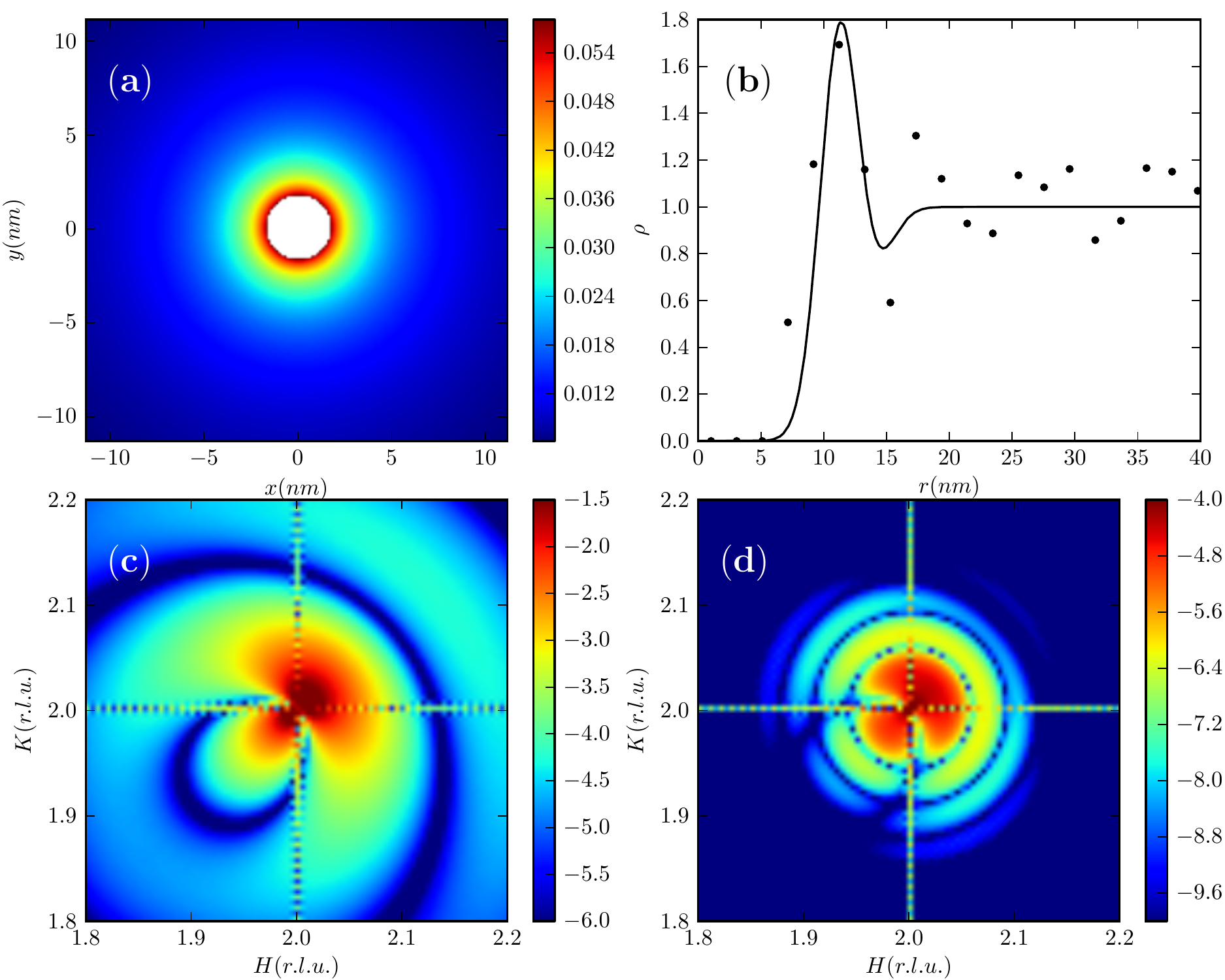}
	\caption{\label{fig:AnalyticSimu} (color online) (a) Amplitude of the atoms radial displacements (in nm), following an isotropic elastic model around a cylindrical inclusion.(b) Radial pair distribution function of the nanocolumns, measured experimentally (dots) and fitted using a gaussian + error function model. (c) Scattering from a single isolated nanocolumn around the $(220)$ reflection, including the scattering contributions from the displaced atoms as well as the missing atoms inside the column (see text for details). (d) Scattering of an assembly of nanocolumns around the $(220)$ reflection, by taking into account their radial distribution function shown in (b). Intensity is plotted with a logarithmic color scale for (c-d).}
\end{figure}

With the further approximation that the columns have a similar size and scatter independently, the scattering is equal to the interference of the scattering of all the columns:
\begin{equation}
A(\vec{k})=\sum_i{A_0(\vec{k})\ e^{2i\pi\vec{k}\cdot\vec{R_i} }}
\end{equation}
where $A_0(\vec{k})$ is the scattering due to a single column and the associated distortion  of the lattice, $\vec{k}$ is the scattering vector and $\vec{R_i}$ is the position of column $i$ in the sample. The intensity can then be written as:
\begin{eqnarray} \label{eq:ScattElasticSimuColumnCorrel}
I(\vec{k})=\vert A(\vec{k})\vert^2=\vert A_0(\vec{k})\vert^2\vert \sum_i{e^{2i\pi\vec{k}\cdot\vec{R_i} }}\vert^2 \nonumber \\ 
=\vert A_0(\vec{k})\vert^2 \times\vert N+2\sum_{i<j}\cos2\pi\vec{k}\cdot(\vec{R_j}-\vec{R_i}) \vert
\end{eqnarray}

With the above approximations the scattered intensity should be proportional to the scattering of a single nanocolumn, multiplied by an interference factor depending on the pair distribution of the columns. If we assume that the inner part of the column is not contributing to the scattering (either disordered or amorphous), then the scattering of the single column (again calculated as a difference with the scattering from a perfect infinite lattice) is the sum of the contribution from the displaced atoms in the Ge matrix and from the `hole' left in the Ge matrix by the column:
\begin{eqnarray} \label{eq:eqDisplHole}
A_0(\vec{k})= A(displaced\ atoms) + A(hole)\nonumber \\
=\sum_i^{displ. at.}f_i\ e^{2i\pi\vec{k}\cdot\vec{r_i^0}}\ (e^{2i\pi\vec{k}\cdot\vec{u_i}}-1)-\sum_j^{hole}f_j\ e^{2i\pi\vec{k}\cdot\vec{r_j^0}}
\end{eqnarray}
where $\vec{r_i^0}$ denotes the original position of atom $i$, and $\vec{u_i}$ its displacement. 

The so-called Huang scattering \cite{huang_x-ray_1947,dederichs_diffuse_1971,pietsch_high-resolution_2004} typically yields nodal planes where the intensity is zero - which in this case would be expected for a plane perpendicular to the scattering vector. However this would only be true if only the $A(displaced\ atoms)$ term was present - the contribution from the hole makes the nodal plane disappear.

We used the following analytical model\cite{hasegawa_stress_1992} for the radial displacements around a cylindrical nanocolumn:

%

\begin{equation}
u(r>R)=\frac{\delta(1+\nu)}{4\pi r^2(1-\nu)}  \\
\end{equation}

where R is the column's diameter, $\delta$ is a parameter giving the extent of the deformation of the matrix, and $\nu$ is the average Poisson coefficient for germanium. For the sake of simplicity we did not take into account the elastic anisotropy of germanium\cite{wortman_youngs_1965} in this model.

In the next section, we use atomic positions generated by this model to compute the scattering around Bragg reflections.

\subsection{Calculated X-ray scattering from the analytical model}

Grazing incidence X-ray scattering was calculated using the atomic positions calculated from the elastic model: the displacements from an ideal germanium lattice are shown in Fig.\ref{fig:AnalyticSimu}a. To take into account the pair distribution of columns, we measured the pair distribution using TEM plane view (Fig.\ref{fig:FFT}a), and modeled it using an isotropic radial distribution function, as can be seen in Fig.\ref{fig:AnalyticSimu}b. As the sample exhibits no structure above the surface (embedded nanocolumns), transmission coefficients \cite{dosch_critical_1992} for the incoming and outgoing X-ray beams at the surface  were not taken into account, as they only affect the scattering as a constant scale factor; the intensity was therefore directly calculated using Eq.\ref{eq:ScattElasticSimuColumnCorrel}.

The results of the simulation around a $(220)$ reflection are shown in Fig.\ref{fig:AnalyticSimu}c for a single isolated column, and in Fig.\ref{fig:AnalyticSimu}d when taking into account pair correlations. In these simulations the intensity is larger at high Bragg angle, which is due to the compression of the Ge matrix by the nanocolumn. The pair correlation induces the correlation ring that is observed experimentally.

However the experimentally observed intensity still differs quantitatively from the simulated one, which is due to the ideal modeling of the lattice: (i) the columns are assumed to be identical (whereas the column diameter distribution is a Gaussian with an FWHM of almost $\approx30\%$), (ii) the radial distribution will also differ depending on the column size, (iii) the true deformation of the Ge matrix should take into account the elastic anisotropy of germanium,\cite{wortman_youngs_1965} and (iv) the distance between columns (nearest neighbor distance of $\approx8\ nm$) is such that the displacement fields generated by all columns will not be independent. For this reason it is particularly interesting to perform atomistic simulations to obtain more realistic atomic positions taking into account inter-column interactions, which is presented in the next section.

\begin{figure} \centering \includegraphics[width=0.48\textwidth]{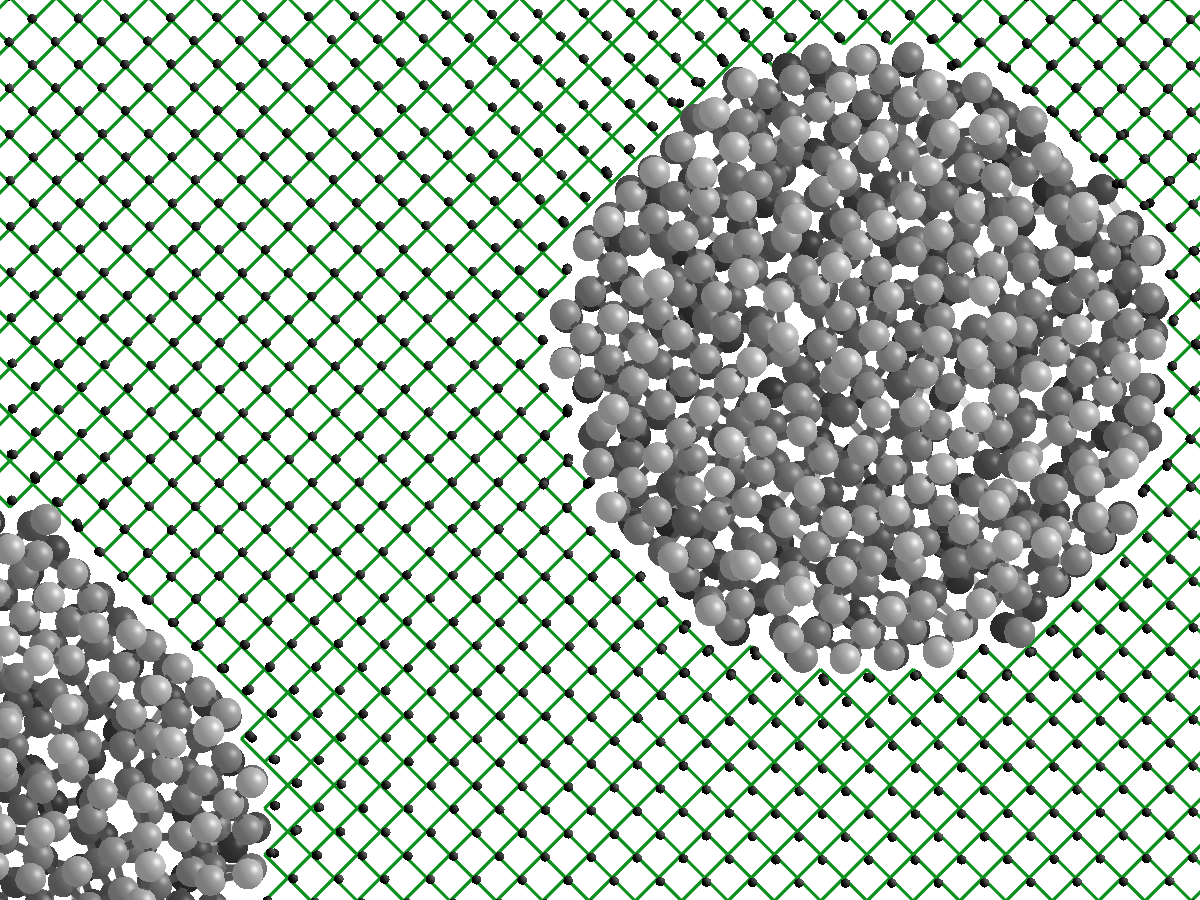}
   \caption{\label{fig:atomicmodel} (color online) Atomic model obtained using a Stillinger-Weber potential, combining amorphous columns (grey spheres; darker shade indicates larger depth) surrounded by the germanium matrix (Ge atoms as black disks). Ge-Ge bonds of the perfect (undistorted) lattice are shown in green. Note the atomic displacements of the Ge atoms in the matrix, due to the column expansion.}
\end{figure}

\subsection{Atomistic simulations}

In order to get a realistic elastic displacement field of the germanium matrix, atomistic simulations were conducted using a Stillinger-Weber potential \cite{stillinger_computer_1985,ding_molecular-dynamics_1986} for the atomic interactions -- one advantage of this potential being that it was developed for both ordered and disordered condensed matter phases.
To simulate a collection of several columns, we used a domain consisting of 100$\times$100 Ge-diamond unit cells, while the size along the column axis was 4 unit cells. Periodic boundary conditions were used in the three space directions.
The column density was chosen to be the experimental value, corresponding to an average of 42 columns per simulation cell (13$\,$100 $\mu$m$^{-2}$).

\begin{figure} 
	\centering
	\includegraphics[width=0.48\textwidth]{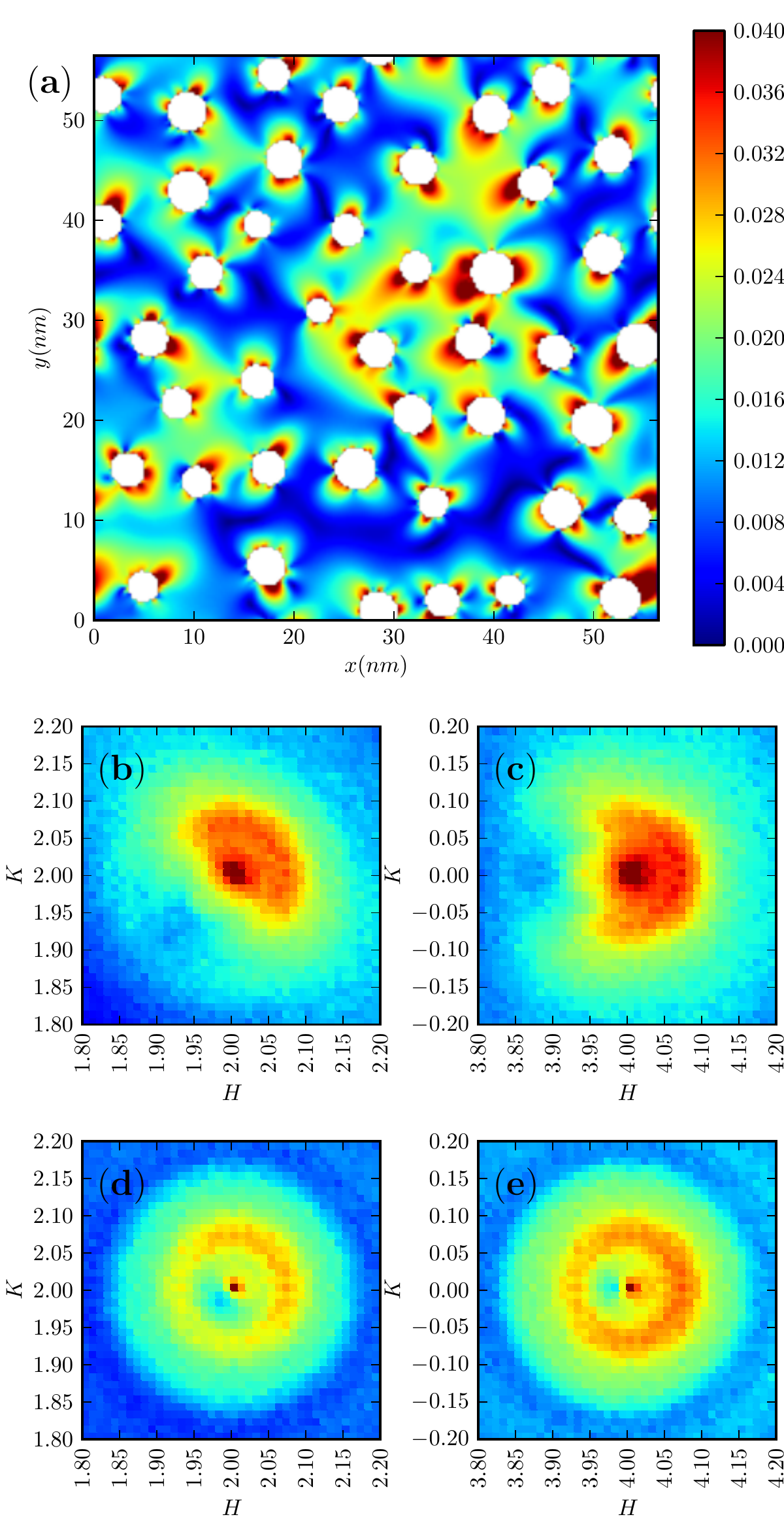}
	\caption{\label{fig:CartoDiffractionSimu} (color online) (a) Mapping of the simulated atomic displacements (from the reference perfect Ge lattice, displacements are given in $nm$), for a simulation with 100x100 unit cells and 42 GeMn nanocolumns (see text for details). Reciprocal space maps have been calculated by summing the intensities from 20 independent atomistic simulations, to lower speckle effects. The maps around (b) the (220) and (c) the (400) reflection exhibit the same features as in Fig.\ref{fig:CartoDiffraction}(d,e). In (d) and (e) are presented the same maps as for (b) and (c), but where atomic displacements have been reduced by a factor 5. The ring observed around the Bragg reflections is located at the position determined by the average distance between nanocolumns, and the polarization of its intensity (high \textit{vs} low angle) is determined by the amplitude of the atomic displacements of the atoms in the matrix surrounding the columns. Intensity is plotted with a logarithmic color scale for (b-e), with a full scale corresponding to 4 orders of magnitude.}
\end{figure}

Starting from an empty simulation cell, the column's centers were added successively at random positions but with an exclusion distance between each new column and the previous ones.
This exclusion distance had a Gaussian distribution with a mean equal to 6.9~nm and a standard deviation  equal to 0.35~nm. The resulting pair correlation has a peak at 7.6~nm close to the experimental value.
The diameters of the columns were also set with a Gaussian distribution of mean and standard deviation respectively equal to 3.6~nm and 0.5~nm (corresponding to a FWHM of $\approx$ 1.2~nm).
Ge atoms were set on the germanium diamond lattice outside of the columns.

Because we are primarily interested in this article by the matrix deformation and not by the precise structure of the GeMn columns themselves, we set atoms also interacting with the Stillinger-Weber potential, but at random positions inside the columns.
Their initial density was the germanium density, but the lower density of the 
resulting amorphous structure leads to nanocolumns compressing the Ge matrix, as 
can be seen in Fig.\ref{fig:atomicmodel}.

A standard minimization of the total potential energy was then performed by conjugate gradient method.
The resulting structure was a collection of amorphous nanocolumns embedded in a strained germanium lattice, as can be seen in Figs. \ref{fig:atomicmodel} and \ref{fig:CartoDiffractionSimu}a.

To be able to simulate the \textit{incoherent} diffraction from separate areas of a sample, we generated in the same way 20 different domains (with the same statistical properties) of $100\times100$ unit cells.

\subsection{Calculated X-ray scattering from the atomistic simulations}

Grazing incidence X-ray scattering was calculated using the atomistic simulations: for the same reasons as before, transmission coefficients for the incoming and outgoing X-ray beams were not taken into account. The intensities were summed for all 20 independent domains of $100\times100$ unit cells, in order to average any `speckle' structure that could arise from the absolute configuration of the 42 nanocolumns in each simulation. The result of these simulations are shown in Fig.\ref{fig:CartoDiffractionSimu}(b-c) for the $(220)$ and $(400)$ reflections, which compare well with the observed maps in Fig.\ref{fig:CartoDiffraction}(b-c).

The main difference with the results obtained using an elastic model (section \ref{sec:AnalyticalModel}) is that there are fewer correlation rings (only the first order is now visible), which is both due to the distribution of columns diameters in the atomistic simulation as well as atomic displacements which correctly take into account the influence of neighboring columns.

In order to investigate the \textit{quantitative} influence of the deformation of the Ge matrix on the asymmetry of the intensity in the correlation rings, we modified the simulated atomic positions by decreasing the displacement (with respect to an undistorted Ge lattice) of all atoms by a factor between 0 and 1. This approach is possible since atomic displacements are, in the elastic regime, proportional to the misfit introduced by the nanocolumns. The RSM maps for a factor equal to $0.2$ is shown in Fig.\ref{fig:CartoDiffractionSimu}(d-e), and reproduce well the rings observed for the sample with the oxidized columns in Fig.\ref{fig:CartoDiffraction}(f-g).

\begin{figure} \centering \includegraphics[width=0.4\textwidth]{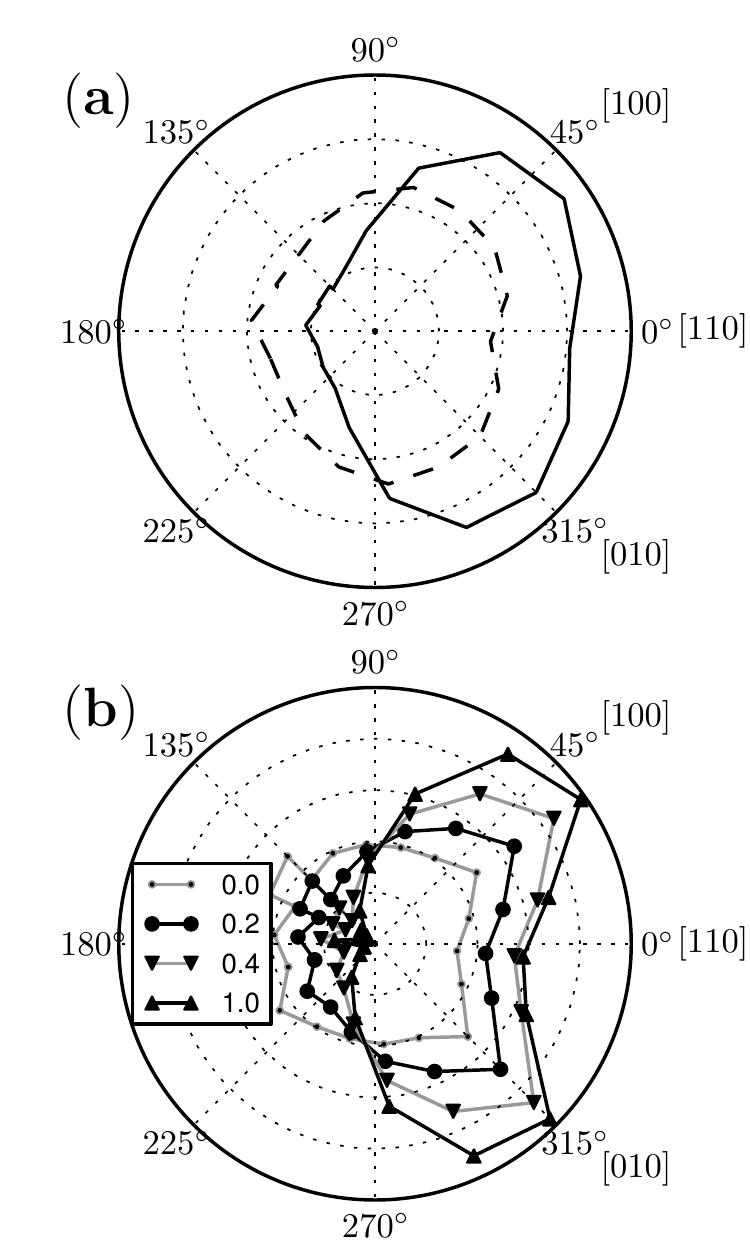}
	\caption{\label{fig:AngularDistribution} Normalized angular distribution of the intensity on the ring surrounding the $(220)$ reflection. This distribution is characteristic of the amplitude of deformation of the matrix lattice around the nanocolumns,  with a larger asymmetry in the radial direction ($0\ vs\ 180^{\circ}$) for larger deformations: $(a)$ distribution observed for a sample grown in situ (continuous line) and another (dashed line) with oxidized columns; $(b)$ simulated distribution of intensities, using a single nanocolumn, for various amplitudes of deformations - the relative displacement amplitudes around the columns are given in the inset - for the largest amplitude the maximum displacement is $\approx0.05\ nm$ for atoms near the border of the columns.}
\end{figure}

A more quantitative comparison of the angular distribution of the intensity is presented in Fig.\ref{fig:AngularDistribution}: the amplitude is maximum at $\pm45^\circ$ with respect to the scattering vector, \textit{i.e.} along the $[100]$ and $[010]$ directions around the $(220)$ reflection, both in the simulation and in the experimental data. 

In Fig.\ref{fig:AngularDistribution}b, the angular distribution is shown as the amplitude of the atomic displacements (with respect to the perfect Ge lattice) is reduced, from a maximum displacement of $\approx0.05\ nm$ down to no displacement: there is a clear reduction of the asymmetry in the angular distribution, which can be used as an indication of the amplitude of the Ge matrix strain around the columns. 

Note that the oxidizing mechanism occurring in the nanocolumns is not known here - obviously as the stress diminishes, the atomic density in the nanocolumns should also decrease, which probably indicates a migration of some atoms of the columns to be replaced by oxygens. Also note that during this process, the diameter of the (oxidized) columns may have increased: but given the width of the size distribution of the nanocolumns (FWHM $\approx$ 1.2~nm), this would only moderately affect the scattered amplitude. Therefore, the decrease in the angular asymmetry of the intensity can directly be linked to the decrease of the strain in the Ge matrix.

\section{\label{sec:conclusion} Conclusion and summary}
In this work we have shown that grazing incidence X-ray scattering maps can be used for a quantitative analysis of layers with GeMn nanocolumns embedded in germanium. In these samples the density of nanocolumns -- related to the spinodal decomposition mechanism and the overall manganese concentration -- leads to short ($\approx8\ nm$) correlation distances, which can directly be measured from X-ray scattering maps under the form of correlation rings around Bragg peaks of the germanium matrix. Moreover the angular intensity distribution around these Bragg peaks shows a direct relation with the amplitude of the deformation in the matrix, which can range up to $\approx0.05\ nm$ and is found to decrease upon oxidation of the columns. A more systematic study of the effect of aging in GeMn nanocolumns (including transmission electronic measurements and SQUID magnetometry) is under way to fully understand the oxidation process and its effect on magnetic properties.

Finally, while this article was focused on the study on the correlations between nanocolumns and the strain of the Ge matrix, several features indicate that there is a contribution for the inner part of the columns, such as the diffuse scattering between the $(220)$ and $(040)$ reflections as well as the weak, large scattering around the forbidden $(130)$ reflection, as can be seen in Fig.\ref{fig:CartoDiffraction}. This aspect of the structure of GeMn layers will be developed in a separate article, notably considering possible atomic structures within the nanocolumns.\cite{rovezzi_atomic_2008,arras_e._metastability_2010}

\acknowledgements
The authors would like to thank J.-S. Micha, T. Sch\"ulli, N. Boudet, G. Carbone and J. Eymery for their help during the experiments. Till Metzger is acknowledged for suggesting the original grazing incidence experiments. The beamlines ID01, CRG BM02 and BM32 of the ESRF are acknowledged for providing beamtime. 

\bibliography{Article-GeMn}
\end{document}